\documentclass{appolb}

\usepackage[title,titletoc]{appendix}
\usepackage[utf8]{inputenc}
\usepackage[square,sort,comma,numbers]{natbib}
\usepackage{graphicx}
\usepackage{lineno}
\usepackage[breaklinks=true]{hyperref}
\usepackage{hypernat}
\usepackage{epsfig}
\usepackage{wrapfig}

\usepackage{graphicx,type1cm,eso-pic,color}

\usepackage{fancyhdr}
\usepackage{pslatex}   
\usepackage{color}
\usepackage{amsmath, amsthm, amssymb} 
\usepackage{endnotes}
\usepackage{xspace}
\usepackage{multirow}

\newcommand {\kt} {\mbox{$k_{\rm T}$}~}

\begin{document}
\title{Extracting femtoscopic radii in the presence of significant additional correlation sources%
%\thanks{Presented at ...}%
}
\author{\L{}. K. Graczykowski, A. Kisiel, M. A. Janik, and P. Karczmarczyk
\address{Faculty of Physics, Warsaw University of Technology, ul. Koszykowa 75, 00-662 Warsaw, Poland}
\\
%{Third Author of different affiliation
%}
%the Name(s) of other Author(s)
%\address{affiliation}
}
\maketitle
%\linenumbers
\begin{abstract}
The Large Hadron Collider has provided large amounts of data on collisions of small systems, such as proton--proton and proton--lead at unprecedented collision energies. Their space-time size and structure can be inferred from the measurement of the femtoscopic correlations for pairs of identical particles. The analysis is complicated by the presence of significant additional sources of two-particle correlations, which influence the correlation function in the region of the femtoscopic effect. In this work we use p--Pb events generated in a model that includes such additional correlation sources to characterize them and propose a robust method of taking them into account in the extraction of the femtoscopic information.  
\end{abstract}
  
\section{Introduction}
\label{sec:intro}

The Large Hadron Collider (LHC) operating at CERN has delivered high statistics data on collisions of small systems during the LHC run 1. The pp collisions have been recorded at $\sqrt{s}=$~0.9, 2.36, 2.76, 7, and 8~TeV. At the end of the LHC run 1, the p--Pb collisions at $\sqrt{s_{\rm NN}}=$~5.02~TeV have been registered. The data analysis for those systems can be treated as an important baseline for the heavy-ion collision studies. In particular, the Pb--Pb collisions at $\sqrt{s_{\rm NN}}=$~2.76~TeV at the LHC reveal that the system created there behaves collectively and undergoes a transition to the deconfined state, the Quark-Gluon Plasma (QGP)~\cite{Aamodt:2010jd,Chatrchyan:2011sx,Aad:2010bu,Aamodt:2010pa,ALICE:2011ab}. One of the observables necessary to draw such conclusions is the femtoscopic measurement for two identical particles (sometimes called
the Bose-Einstein correlation, or HBT measurement)~\cite{Lednicky:2005af,Lisa:2005dd}. It allows to extract the size of the emitting source and study its dynamics, as well as to constrain the models attempting to describe such systems. At first the collective effects were not expected in the reference systems, such as pp and p--Pb. However, some recent experimental results~\cite{Abelev:2012ola} suggest that a collective system might be created in the high multiplicity p--Pb collisions and corresponding calculations from the hydrodynamic models support such claim~\cite{Bozek:2012gr,Bozek:2013uha,Qin:2013bha}. Alternative predictions were provided by models based on the Color-Glass Condensate formalism where the system size in pp and p--Pb collisions are expected to be similar~\cite{Dusling:2012wy,Dusling:2013oia,Bzdak:2013zma}. The various predictions of system size differ from each other by 30-50\%. Therefore, answers to important physics questions rely on a precision measurement of femtoscopic sizes in small systems. It is though of high importance that a reliable method to measure such sizes is proposed for such collision systems. 

The femtoscopic correlation is measured as a function of the relative momentum $\mathbf{q}=\mathbf{p_1}-\mathbf{p_2}$, most often for the pairs of identical pions. Due to the fact that pions follow the Bose-Einstein statistics, a significant positive correlation is observed as $q=|\mathbf{q}|$ nears 0. The width of this enhancement is inversely proportional to the system size. In the ideal case such correlation rests on a flat baseline, reflecting the lack of other two-particle correlations. Such scenarios is indeed realized for example for heavy-ion collisions, where all other correlations are either small or have a $q$ scale vastly different than the femtoscopic effect. This is not the case for collisions in small systems, where a relatively small number of particles is produced. Measurements done by various experiments~\cite{Chajecki:2009zg,Aggarwal:2010aa,Aamodt:2010jj,Aamodt:2011kd,Abelev:2014pja,Adare:2014vri,Chajecki:2005qm,Khachatryan:2010un,Khachatryan:2011hi} show that significant additional correlation sources are contributing to the two particle correlation function. We will later collectively refer to such effects as "non-femtoscopic" background, as we are primarily interested in extracting the femtoscopic signal. These correlations have a magnitude and width in $q$ comparable to the femtoscopic signal, and therefore the two cannot be easily disentangled. The sources of such correlations are, among others, the energy-momentum conservation and the "mini-jet" phenomena~\cite{Aamodt:2011kd}.

Two main approaches have been taken by experiments to deal with the "non-femtoscopic" correlations, both relying on the modelling of the background in Monte Carlo (MC) models. The first is to construct a "double-ratio", where the experimental correlation function is divided by a corresponding one from the MC calculation. This technique relies on the fact that the particle production process in MC does not take into account the Bose-Einstein enhancement, but it does include other sources of correlation. The application of the "double-ratio" technique should be equivalent to "dividing out" the non-femtoscopic effects and leave a pure Bose-Einstein signal. The second technique is to parametrize the background in MC calculation and then use it as an additional term in the fitting function applied to the experimental correlation function. We note that both approaches are, in perfect conditions (small size of $q$ bins in the correlation function, large statistics both for data and MC, etc.), mathematically equivalent. However, the method with the additional term in the fitting function offers greater flexibility, which is needed for this work.

In this work we perform a methodological verification of the procedures used to account for the "non-femtoscopic" background. Using the EPOS 3.076 model~\cite{Werner:2013tya,Werner:2013ipa}, we calculate the three-dimensional correlation functions in the Longitudinally Co-Moving System (LCMS)~\cite{Bertsch:1989vn,Pratt:1986cc}, where the pair momentum along the beam vanishes. The correlation functions are calculated with (1) pure Bose-Einstein signal, (2) with the background effects only, and (3) with both correlation sources combined. We extract the source size from the "pure" correlations functions, and compare them with the ones extracted from the "full" calculation, where the background is constrained using the "only background" calculation. We propose several methods to parametrize the background. We estimate the systematic uncertainty coming from their
application and discuss their validity and stability. We conclude by selecting the method which is most reliable and introduces the smallest uncertainty in the procedure.

\section{Description of the Monte Carlo simulations}
\label{sec:model}

\subsection{Choice of model}
\label{sec:eposintro}

In order to perform the calculations planned for this work, a Monte Carlo event generator able to perform a calculation for a small system, such as p--Pb (including realistic modelling of minimum-bias collisions with "mini-jet" effects), is required. The model must also provide, for each particle, information crucial for femtoscopy, such as freeze-out coordinates. The EPOS ver. 3.076~\cite{Werner:2013tya,Werner:2013ipa} was chosen and run with the parameters corresponding to the p--Pb collisions at $\sqrt{s_{\rm NN}}=$~5.02 TeV, the same as the recent p--Pb run at the LHC. The
model is based on the Regge formalism and includes fragmentation of partons scattered with moderate energy which are usually associated with the "mini-jet" phenomena. As such, it produces significant non-femtoscopic correlations, which are in qualitative agreement with the trends observed in data. The minimum-bias sample of events was generated, containing all information about the produced particles, including their freeze-out coordinates.

\subsection{Femtoscopic formalism}
\label{sec:femtoformalism}
 
The femtoscopic correlation function is, by definition, a ratio of the conditional probability to observe two particles together, normalized to the product of probabilities to observe each of them separately. Experimentally it is measured by dividing the distribution of relative momentum of pairs of particles detected in the same collision (event) by an equivalent distribution for pairs where each particle is taken from a different
collision, usually using the "event mixing" technique. Femtoscopy focuses on the mutual two-particle correlation which comes from the (anti-)symmetrization of the wave function for pairs of identical particles. Another source of correlation is the Final State Interaction (FSI),
that is Coulomb or strong. At the moment no MC models exist that would take the effects of two-particle wave-function symmetrization or the FSI into
account when simulating particle production. The effect is usually added in an "afterburner" code, which requires the knowledge of each particles' emission point and momentum (this information must be provided by the MC model). We employed this procedure in this work. 

The femtoscopic correlation function can be expressed as: 
\begin{equation}
C(\mathbf{q}) = \int S({r}, \mathbf{q}) |\Psi(\mathbf{q},r)|^{2}
d^{4}r,
\label{eq:cfrompsi}
\end{equation}
where $r$ is a relative space-time separation four-vector of the two particles and $S$ is the source emission function which can be interpreted as a probability to emit a given particle pair from a given set of emission points with given momenta. For pairs of identical charged pions, which we consider in this work, the pair wave function must be symmetrized. In addition, pions also interact via the Coulomb and strong interaction. However, strong interaction in this case is expected to be small~\cite{Lednicky:2005tb}, therefore we limit our considerations to Coulomb interaction only. The mutual final-state interaction is then reflected in the pair wave-function\footnote{More precisely it is the Bethe-Salpeter amplitude for the pair, corresponding to the solution of the quantum scattering problem taken with the inverse time direction} giving~\cite{Lednicky:2005tb}:  
\begin{equation}
%\Psi^{(+)}_{-k^{*}}({\bf r^{*}}) = \sqrt{A_{\rm C} (\eta)} \left [ e^{-i
%\vec k^{*} {\bf r^{*}}} F(-i \eta, 1, i \zeta) + f_{\rm C} (\vec k^{*})
%\frac {\tilde G (\rho, \eta)} {{\bf r^{*}}} \right ]
\Psi^{(+)}_{-\mathbf{k^{*}}}({\bf r^{*}}) = \sqrt{A_{\rm C} (\eta)}
\frac{1}{\sqrt{2}} \left [ e^{-i \mathbf{ k^{*}} {\bf r^{*}}} F(-i \eta, 1,
  i \zeta^{+}) + e^{i \mathbf{ k^{*}} {\bf r^{*}}} F(-i \eta, 1,
  i \zeta^{-}) \right ],
\label{eq:fullpsi}
\end{equation}
where $A_{\rm C}$ is the Gamow factor, $\zeta^{\pm} = k^{*} r^{*} (1 \pm \cos{\theta^{*}})$, $\eta = 1/(k^{*} a_{\rm C})$, and $F$ is the
confluent hypergeometric function. $\theta^{*}$ is the angle between the pair relative momentum $k^{*}=|\mathbf{k^*}|=q/2$ and relative position $r^{*}=|\mathbf{r^{*}}|$ in the Pair Rest Frame (PRF), while $a_{\rm C}$ is the Bohr radius of the pair. Since we are dealing with identical particles, $\Psi$ is properly symmetrized.

\subsection{Calculation of the femtoscopic effect}
\label{sec:afterburner}

To perform the calculation of the correlation function according to Eq.~(\ref{eq:cfrompsi}) with particles produced by the event generator, a Monte Carlo procedure must be applied. First, all charged pions from the EPOS event are combined into pairs. A distribution $B$ is created where each pair is filled with the weight of 1.0, at a corresponding relative momentum $q$. The second distribution $W$ is created, where the pair is inserted in the same manner, but with the weight calculated according to Eq.~(\ref{eq:fullpsi}). To construct the third distribution, two pions in the pair are taken from different EPOS events in a so-called "mixing" technique and pairs are inserted with weight 1.0 in the distribution $M$. Three distinct correlation functions can then be created, each containing a specific set of information. All of them are needed for the study presented in this work. The function $C_{\rm QS}=W/B$ is mathematically equivalent to the Monte Carlo integration of Eq.~(\ref{eq:cfrompsi}). It contains only the "pure" Quantum Statistics + FSI signal. The correlation function $C_{\rm B}=B/M$ contains all the event-wide correlations which are present in the EPOS simulation, including the ones which contribute to the non-femtoscopic effect, but it does not include the QS+FSI correlation. Therefore, it represents the "background" in our study. The third histogram $C_{\rm F}=W/M$ represents the "full" correlation, including both the effects of the QS+FSI, as well as all other correlations contained in the model. The $C_{\rm F}$ most closely resembles an experimental correlation function.

Moreover, since all the distributions are calculated in three dimensions in LCMS, following the approach of the experiments~\cite{Aggarwal:2010aa,Aamodt:2011kd}, we employ a spherical harmonic (SH) decomposition of the measured correlation functions~\cite{Brown:1997ku,Chajecki:2006hn,Kisiel:2009iw}.

\begin{figure}[!ht]
\centering
\begin{minipage}[!ht][][t]{0.483\linewidth}
\includegraphics*[width=1.0\textwidth]{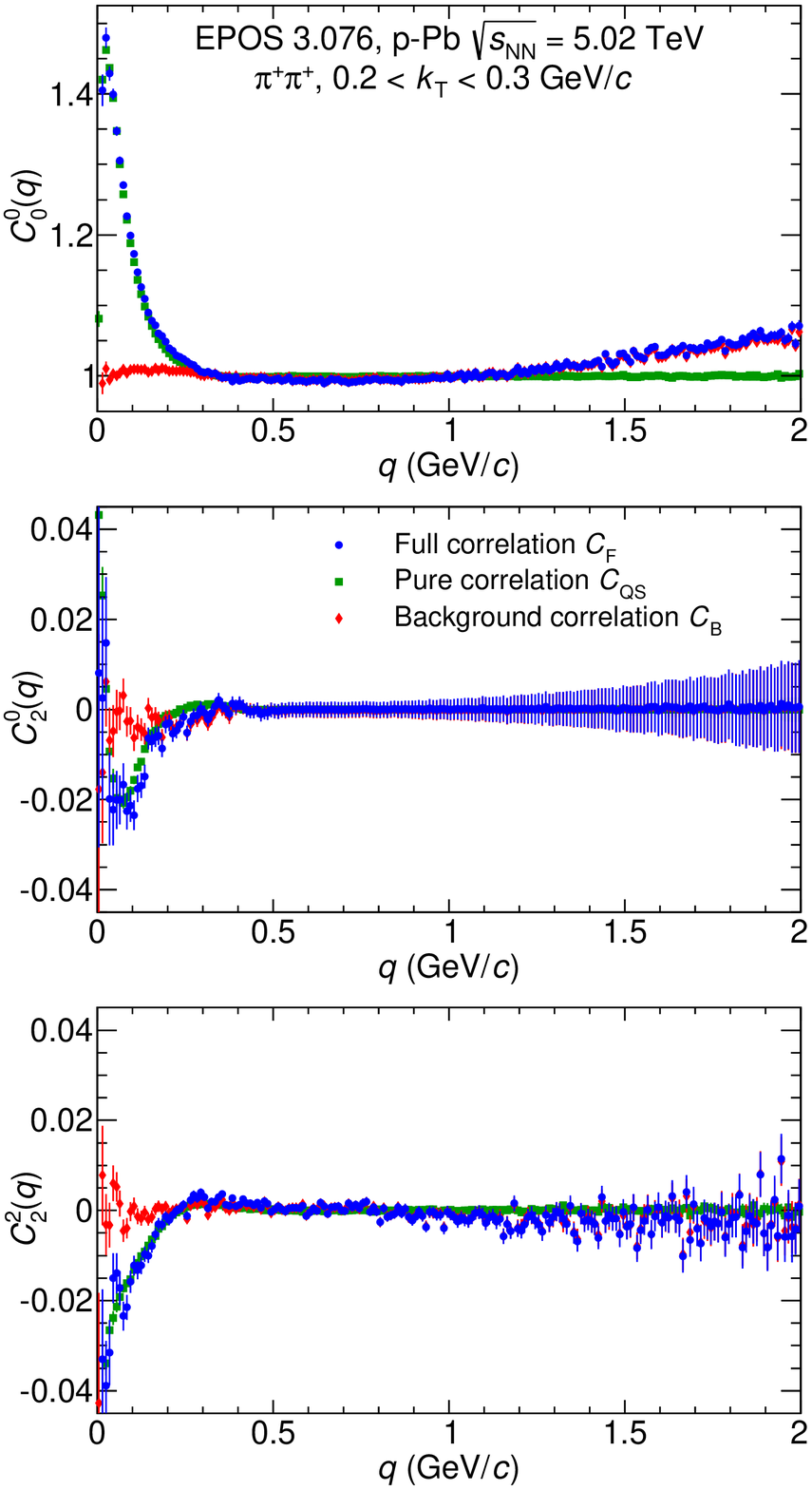} 
\end{minipage}
\hspace{0.25cm}
\begin{minipage}[!ht][][t]{0.483\linewidth}
\includegraphics*[width=1.0\textwidth]{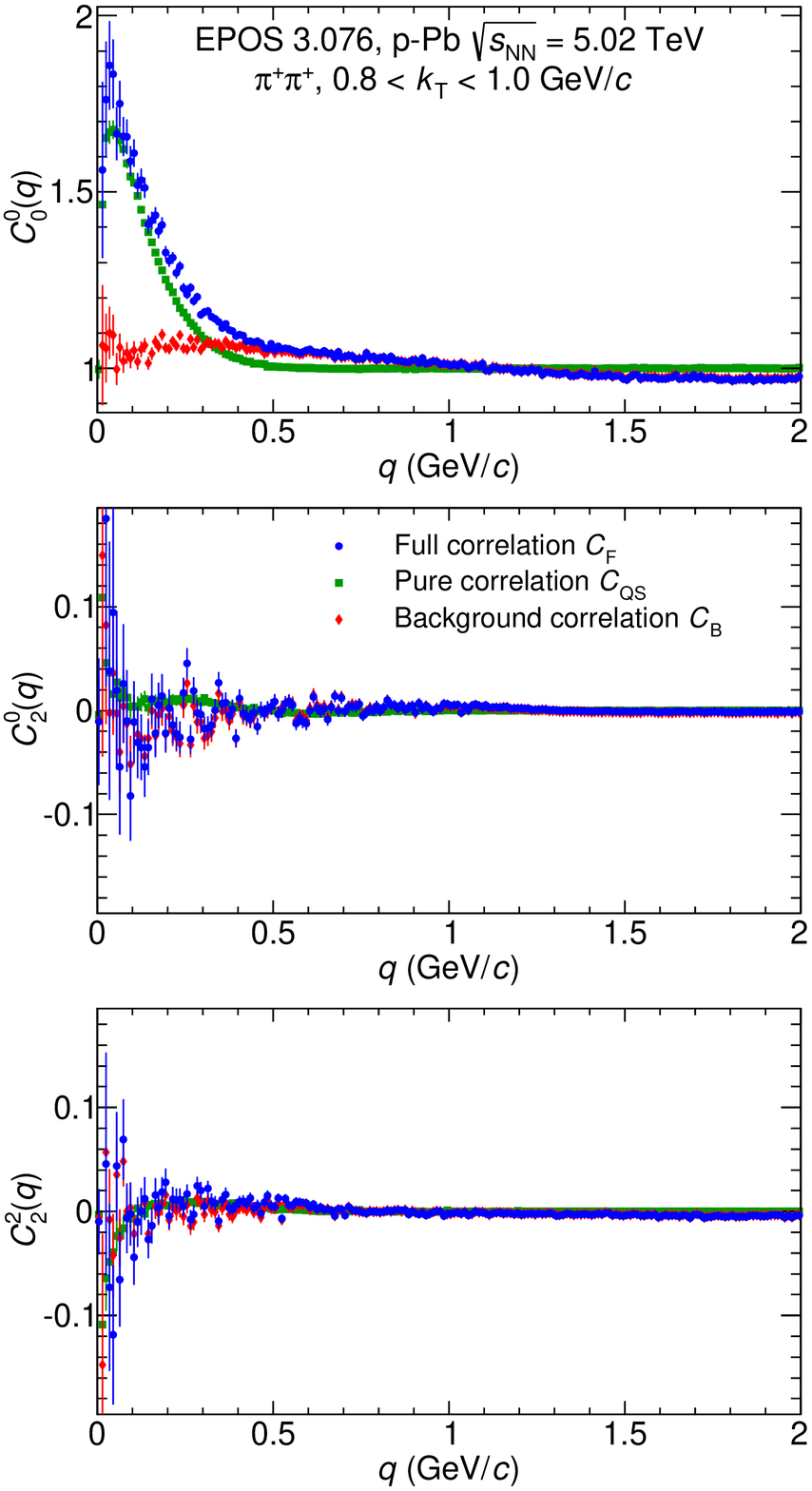}  
\end{minipage}
\caption[]{First three non-vanishing components of the SH representation of the $\pi^+\pi^+$ correlation functions from EPOS model for $0.2<k_{\rm T}<0.3$~GeV/$c$ (left plot) and $0.8<k_{\rm T}<1.0$~GeV/$c$ (right plot).}  
\label{fig:3funs}
\end{figure}

All the correlation functions have been calculated for seven ranges of pair transverse momentum $k_{\rm T}=\frac{|\mathbf{p_{T,1}}+\mathbf{p_{T,2}}|}{2}$: 0.2--0.3, 0.3--0.4,0.4--0.5, 0.5--0.6, 0.6--0.7, 0.7--0.8, and 0.8--1.0~GeV/$c$. An example of all three correlation functions, calculated for two (low and high) \kt ranges are shown in Fig.~\ref{fig:3funs}.

\subsection{Extracting the femtoscopic information}
\label{sec:extractfemto}

With the three correlation functions calculated we proceed to treat them with an experimentalist's recipe. We employ a fitting procedure to extract the femtoscopic radii. In order to derive a fitting function, the functional form of $S$ must be assumed. In heavy-ion collision analysis it is usually assumed to be a three-dimensional ellipsoid with a Gaussian density profile: 
\begin{equation}
S(\mathbf{r}) \approx \exp \left(-\frac{r^{2}_{\mathrm{out}}}
    {4{R^{\mathrm{G}}_{\mathrm{out}}}^2} - \frac{r^{2}_{\mathrm{side}}}
    {4{R^{\mathrm{G}}_{\mathrm{side}}}^2} - \frac{r^{2}_{\mathrm{long}}}
    {4{R^{\mathrm{G}}_{\mathrm{long}}}^2} 
 \right ),
\label{eq:sgauslcms}
\end{equation}
where $r_{\rm out}$, $r_{\rm side}$, and $r_{\rm long}$ are components of $\mathbf{r}$ calculated in LCMS and $R_{\rm out}$, $R_{\rm side}$,
and $R_{\rm long}$ are single-particle femtoscopic source radii. The Coulomb part $K_{\rm C}$ of the charged pion wave-function can be approximately treated as independent from the QS part. It is then integrated separately in a procedure known as Bowler-Sinyukov
fitting~\cite{Bowler:1991vx,Sinyukov:1998fc}. Then Eq.~(\ref{eq:cfrompsi}) gives the following Gaussian fit function (also referred to as "GGG"):
\begin{equation}
C_{\rm qs}(\mathbf{q}) = (1-\lambda) + \lambda K_{\rm C}(q_{\rm inv})
\left [1 + \exp \left(-{R^{\mathrm{G}}_{\mathrm{out}}}^{2}
  q_{\rm out}^{2}-{R^{\mathrm{G}}_{\mathrm{side}}}^{2}q_{\rm side}^{2} -{R^{\mathrm{G}}_{\mathrm{long}}}^{2}q_{\rm long}^{2}
\right) \right ] ,
\label{eq:cfit}
\end{equation}
where $\lambda$ accounts for the fact that not all pion pairs are correlated in the source. $K_{\rm C}(q_{\rm inv})$ is the two-pion Coulomb wave-function integrated on a source with Gaussian density profile. For the source sizes considered here ($<3$ fm) it has significant influence only in a narrow region at small $q$, nevertheless we include the treatment of Coulomb effects in order to be able to follow the experimental procedure as closely as possible. Eq.~(\ref{eq:cfit}) is fitted directly to the calculated correlation functions $C_{\rm QS}$ to extract the "true" femtoscopic radii. 

Experiments reported~\cite{Aamodt:2011kd,Abelev:2014pja,Khachatryan:2011hi} that in small systems the correlation functions deviate significantly from the Gaussian shape given by Eq.~(\ref{eq:cfit}). Therefore, an alternative form, the so-called Exponential-Gaussian-Exponential (or "EGE") was also used and found to describe data better: 
\begin{equation}
C_{\rm qs} (\mathbf{q}) = (1-\lambda) + \lambda K_{\mathrm{C}} \left [
  1 + \exp \left
    (-\sqrt{{R^{\mathrm{E}}_{\mathrm{out}}}^{2}q^{2}_{\mathrm{out}}}-{R^{\mathrm{G}}_{\mathrm{side}}}^{2}q_{\mathrm{side}}^{2}-\sqrt{{R^{\mathrm{E}}_{\mathrm{long}}}^{2}q^{2}_{\mathrm{long}}} \right ) \right ].  
\label{eq:cfitege}
\end{equation}
We use it as an alternative fitting function in this work as well in order to see if a particular shape of the femtoscopic effect influences the background estimation procedure.

In the presence of additional "non-femtoscopic" correlations, the forms given by Eqs.~(\ref{eq:cfit}) or~(\ref{eq:cfitege}) will produce unreliable results. Those effects must be taken into account with additional factors in the fitting equation. Following the discussion in Sec.~\ref{sec:afterburner} such factor should be multiplicative with the QS+FSI effect. A modified fitting function for the Gaussian and the EGE fit case is then:
\begin{equation}
C_{\rm f}(\mathbf{q}) = NC_{\rm qs}(\mathbf{q}) \Omega(\mathbf{q}),
\label{eq:cfitbg}
\end{equation}
where $N$ is the normalization factor and $\Omega$ term contains the "non-femtoscopic" effects. Obviously, the exact form of $\Omega$ is not known. $\Omega$ will also naturally introduce new fitting parameters. The main purpose of this work is to propose a recipe to obtain a form for $\Omega$. We will then apply this procedure to our model calculation and try to extract the "realistic" femtoscopic radii by fitting Eq.~(\ref{eq:cfitbg}) to the calculated $C_{\rm F}$. By comparing these "realistic" radii with the "true" ones obtained from the fit of $C_{\rm QS}$ we will be able to judge the correctness of the procedure to extract $\Omega$ as well as the correctness of the fitting process itself. We will also estimate the theoretical systematic uncertainty coming from the presence of the background.

\subsection{Characterizing the background}
\label{sec:back}
In order to propose a function for the $\Omega$ term needed in Eq.~(\ref{eq:cfitbg}) and accounting for the non-femtoscopic effects in the fitting procedure, we need to calculate $C_{\rm B}$ that contains only non-femtoscopic correlations. Examples of the $C_{\rm B}$ calculated for selected pair \kt ranges are shown in Fig.~\ref{fig:3funs}. The background at low \kt is flat at low $q$, where the femtoscopic effect is most prominent. It
shows a rise at $q>1.0$~GeV/$c$ due to the momentum conservation in "mini-jet" mechanism, however this behavior is not relevant for the femtoscopic analysis. At large \kt there is a significant correlation, wide in $q$, approximately Gaussian in shape, with prominent contribution to the low $q$, where the femtoscopic effect is located. Its three-dimensional shape is reflected in the (2,0) and (2,2) SH components. They differ from zero, but not strongly, indicating that the shape is approximately spherically symmetric in LCMS. 

Fixing the background with the MC calculation introduces a model dependence in the analysis. Therefore we propose several options for the parametrization of $C_{\rm B}$, with varying degree of such model dependence. We propose that the background has a Gaussian shape:
\begin{equation}
\Omega_{0}^{0}(q) = 1+a_{0}^{0} \exp\left (-\frac { q^{2}}
  {2(\sigma_{0}^{0})^{2}} \right),
\label{eq:b00}
\end{equation}
where $a_{0}^{0}$ is a free parameter describing the magnitude of the correlation and $\sigma_{0}^{0}$ is another free parameter describing its
width. In the first scenario, with minimal model dependence, we only fix $\sigma_{0}^{0}$, separately for each \kt range, from the fit to the $C_{\rm B}$. In the second scenario, we fix both the $\sigma_{0}^{0}$, as well as $a_{0}^{0}$ for each \kt range. In the third scenario we also
account for the full three-dimensional shape of the background, with the parametrization of the (2,0) and (2,2) components of the background:
\begin{equation}
\Omega_{2}^{0}(q) = a_{2}^{0}\exp\left (-\frac { q^{2}}
  {2(\sigma_{2}^{0})^{2}} \right)+\beta_2^0,
\label{eq:b20}
\end{equation}
\begin{equation}
\Omega_{2}^{2}(q) = a_{2}^{2}\exp\left (-\frac { q^{2}}
  {2(\sigma_{2}^{2})^{2}} \right) + \gamma_{2}^{2} q+\beta_2^2,
\label{eq:b22}
\end{equation}
where $a_{2}^{0}$, $\sigma_{2}^{0}$, $a_{2}^{2}$, $\sigma_{2}^{2}$, $\gamma_{2}^{2}$, $\beta_2^0$, and $\beta_2^2$ are free parameters of the fit to $C_{\rm B}$. All of them but $\beta_2^0$ and $\beta_2^2$, which are kept free, are then fixed in the fitting of the full correlation function. The overall fitting formula is therefore of the following form:  
\begin{equation}
\begin{split}
C_{\rm f}(\mathbf{q}) = N\cdot C_{\mathrm{qs}}(\mathbf{q})\cdot \left [  \Omega_0^0(q) \cdot Y^0_0(\theta,\varphi)+ \Omega_2^0(q)  \cdot Y_2^0(\theta,\varphi)+ \Omega_2^2(q) \cdot Y_2^2(\theta,\varphi) \right ],
\end{split}
\label{eq:C3D}
\end{equation}
where $Y_0^0(\theta,\varphi)$, $Y_2^0(\theta,\varphi)$, and $Y^2_2(\theta,\varphi)$ are the corresponding spherical harmonics (for definitions see Appendix~\ref{appendix}). 

We stress that this particular functional form has been derived for this particular EPOS MC calculation and is by no means a universal one. Each time such analysis is performed, a new functional form should be proposed, corresponding to the particular background shape observed in data or MC calculation. Nevertheless the three scenarios proposed represent three rather general cases of background characterization. Scenario 1 (also referred to as "Background 1") corresponds to only constraining the background shape in $q$, scenario 2 (also referred to as "Background 2") corresponds to constraining also the background magnitude, while scenario 3 (also referred to as "Background 3") corresponds to fixing the full three-dimensional shape and magnitude of the background. 

\section{Fitting the pure correlation}
\label{sec:fitpure}

\begin{figure}[!ht]
\centering
\begin{minipage}[!ht][][t]{0.483\linewidth}
\includegraphics*[width=1.0\textwidth]{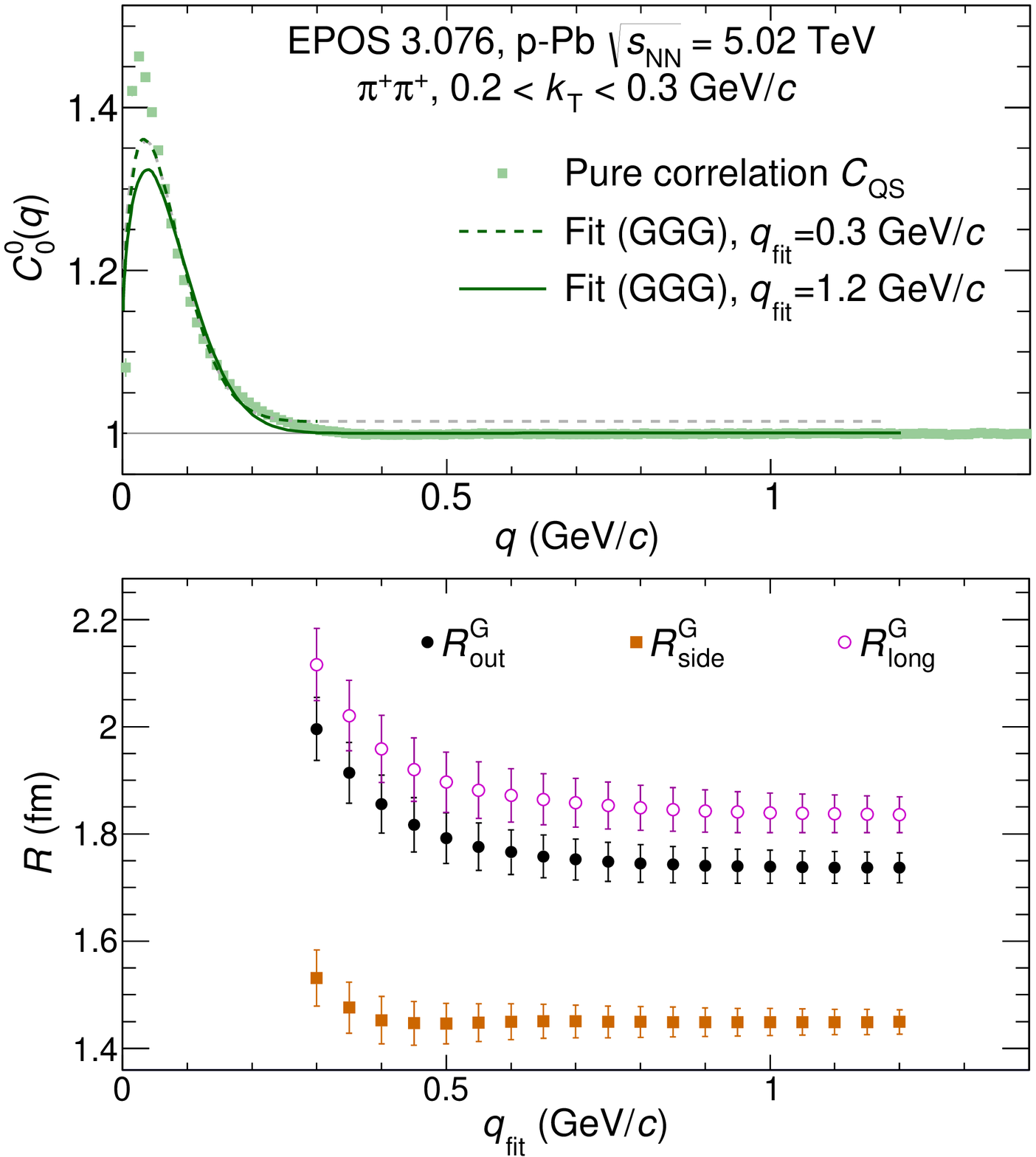} 
\end{minipage}
\hspace{0.25cm}
\begin{minipage}[!ht][][t]{0.483\linewidth}
\includegraphics*[width=1.0\textwidth]{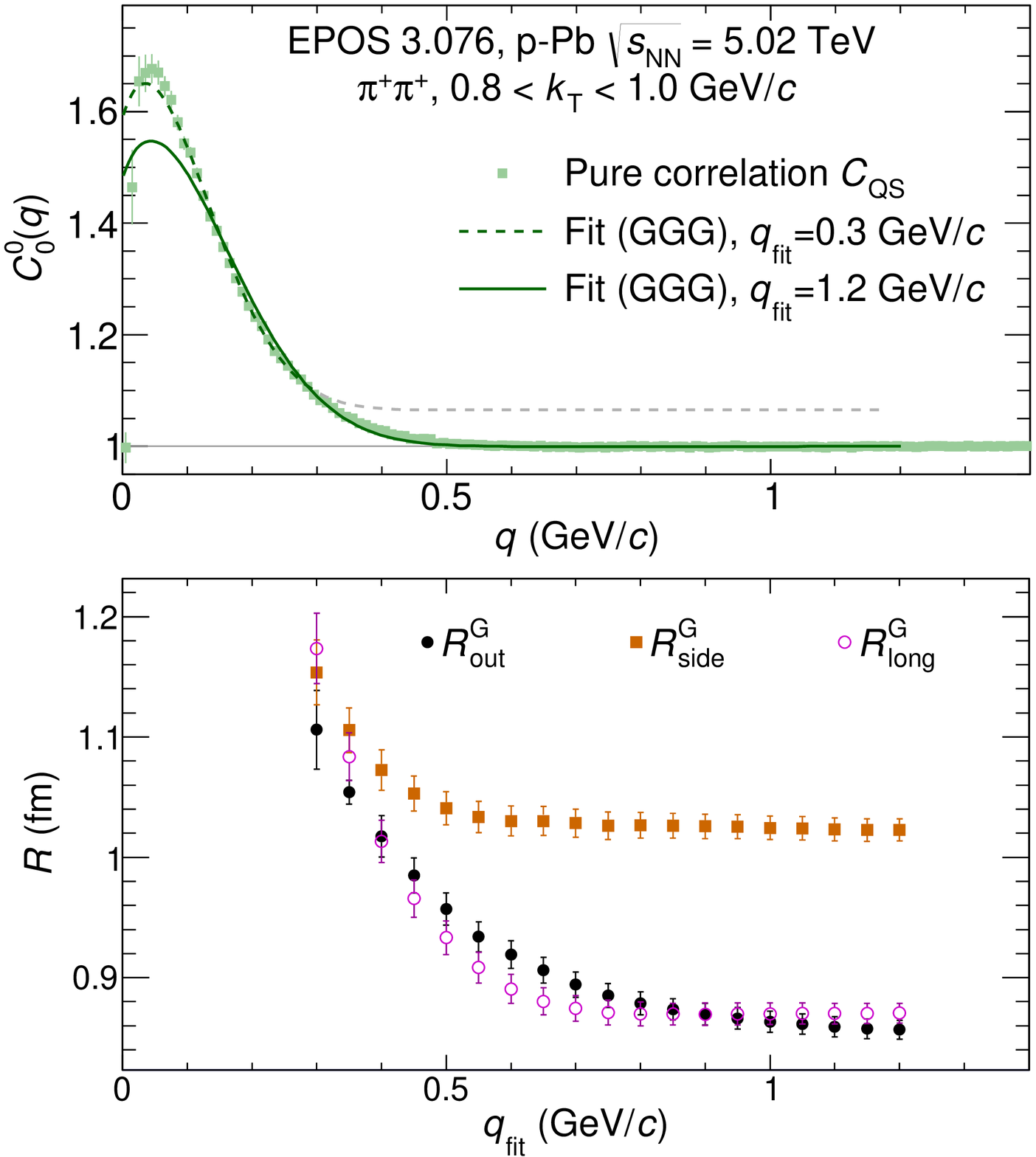}  
\end{minipage}
\caption[]{$C^0_0$ component of the SH representation of pure femtoscopic effect $C_{\rm QS}$ for low and high \kt ranges (upper panels). Extracted Gaussian femtoscopic radii as a function of maximum fit range in $q$ for low and high \kt (lower panels). Dashed and solid lines correspond fo fits with maximum fit range $q_{\rm fit}=0.3$~GeV/$c$ and $q_{\rm fit}=1.2$~GeV/$c$, respectively.} 
\label{fig:fitRangeGGG}
\end{figure}

\begin{figure}[!ht]
\centering
\begin{minipage}[!ht][][t]{0.483\linewidth}
\includegraphics*[width=1.0\textwidth]{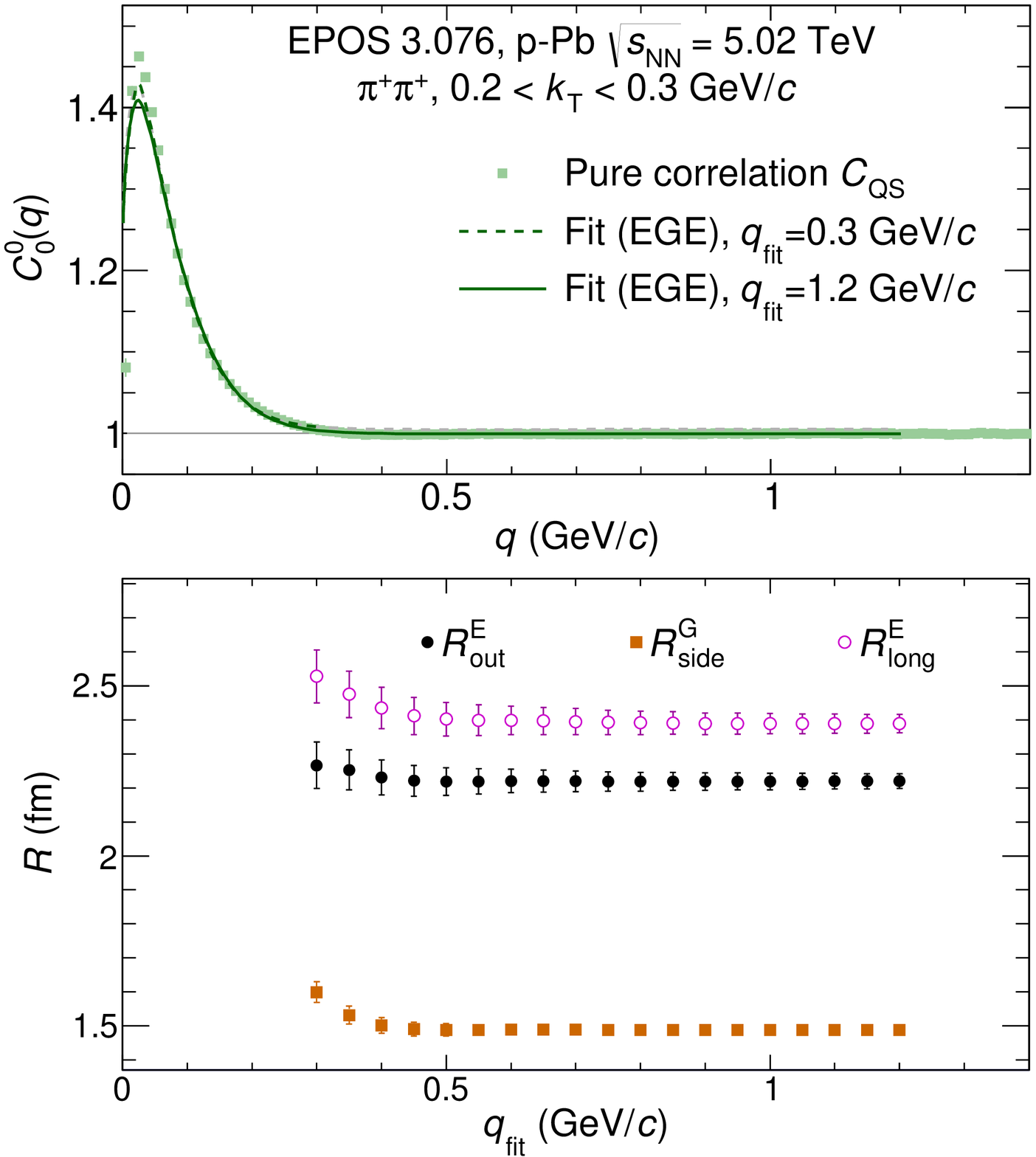} 
\end{minipage}
\hspace{0.25cm}
\begin{minipage}[!ht][][t]{0.483\linewidth}
\includegraphics*[width=1.0\textwidth]{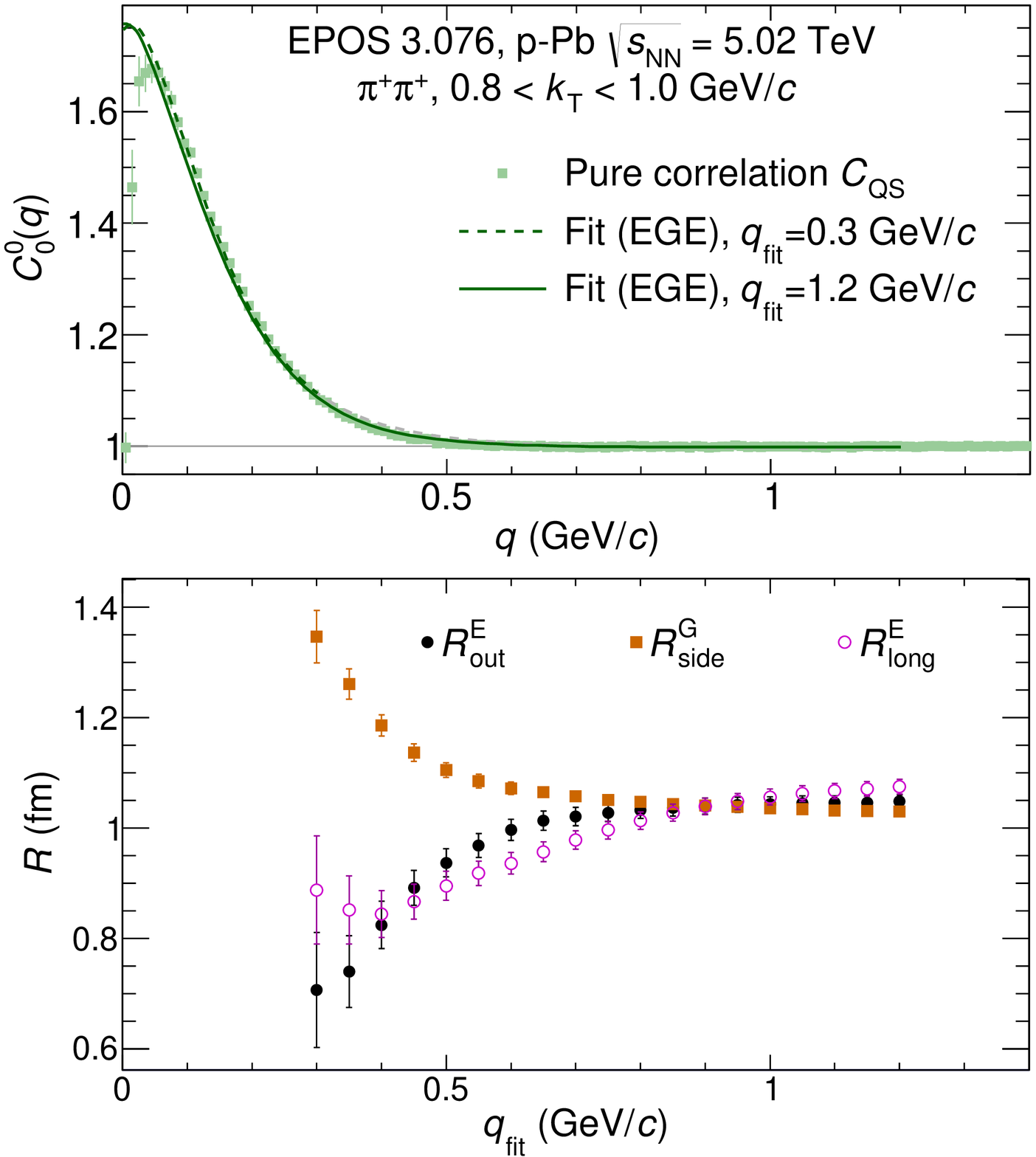}  
\end{minipage}
\caption[]{$C^0_0$ component of the SH representation of pure femtoscopic effect $C_{\rm QS}$ for low and high \kt ranges (upper panels). Extracted EGE femtoscopic radii as a function of maximum fit range in $q$ for low and high \kt (lower panels). Dashed and solid lines correspond fo fits with maximum fit range $q_{\rm fit}=0.3$~GeV/$c$ and $q_{\rm fit}=1.2$~GeV/$c$, respectively.} 
\label{fig:fitRangeEGE}
\end{figure}

The pure correlation function $C_{\rm QS}$ is fitted with formulas given in Eqs.~(\ref{eq:cfit}) and (\ref{eq:cfitege}) to obtain the reference radii. The values of the fit naturally depend on the range in $q$ in which the fit is performed, which is shown in Figs.~\ref{fig:fitRangeGGG} and~\ref{fig:fitRangeEGE}. An expected behavior is seen: when the fitting range is not wide enough, a dependence of the fit parameters on the fitting range is observed. Also for a narrow fitting range the procedure is not able to correctly constrain the normalization of the correlation
function. Both effects are quite pronounced for the Gaussian fits, but they are also present, to a smaller degree, when a more appropriate
shape of the correlation peak, the EGE, is used. Only when the fit range maximum is larger than the width of the femtoscopic effect: around 0.6~GeV/$c$ for the low \kt and 0.8~GeV/$c$ for the high $k_{\rm T}$, the radii values stabilize and do not change further with increase of
the fit range maximum. At the same time the normalization is also properly constrained. 

\begin{figure}[!ht]
\centering
\begin{minipage}[!ht][][t]{0.483\linewidth}
\includegraphics*[width=1.0\textwidth]{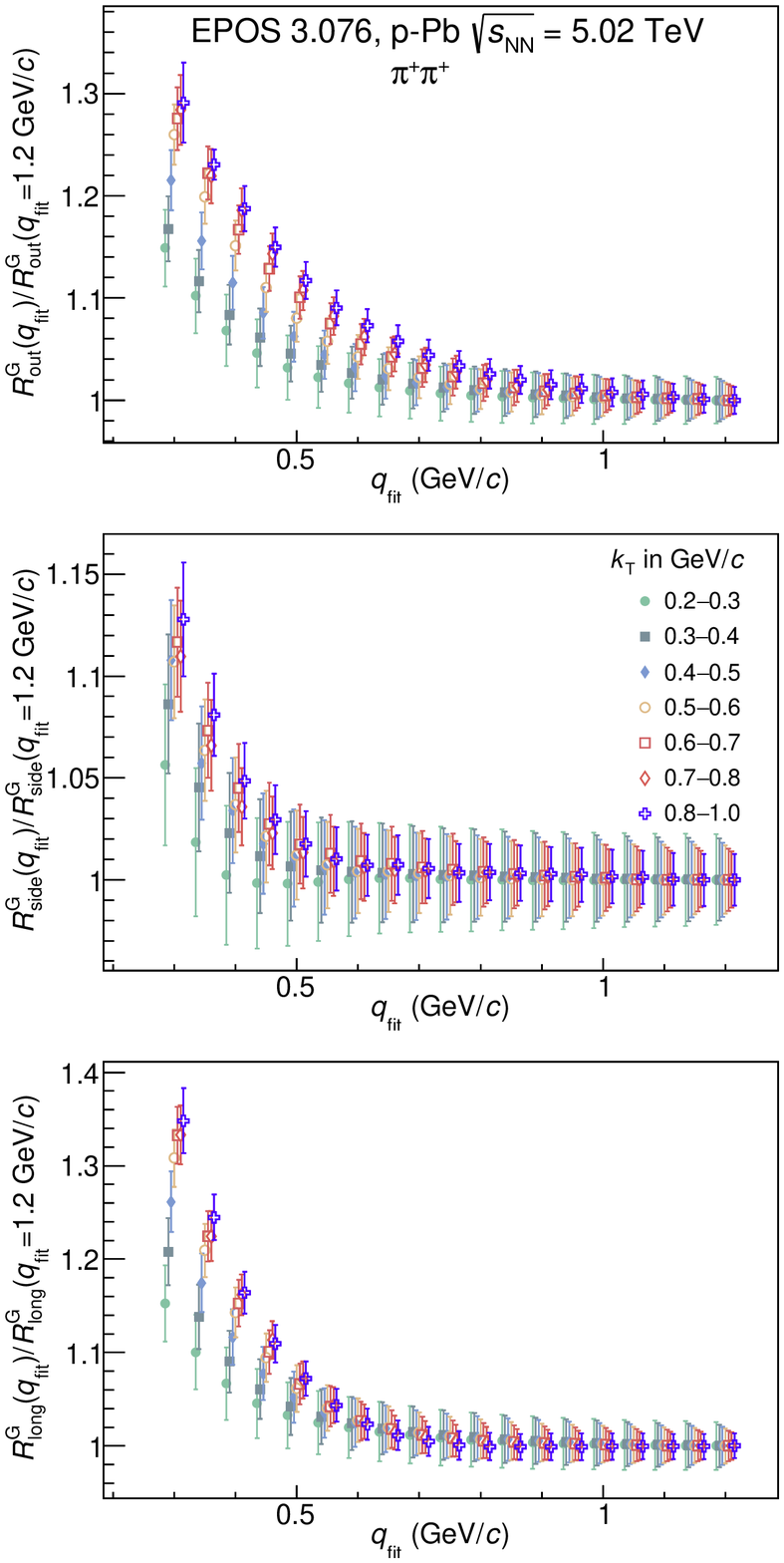} 
\end{minipage}
\hspace{0.25cm}
\begin{minipage}[!ht][][t]{0.483\linewidth}
\includegraphics*[width=1.0\textwidth]{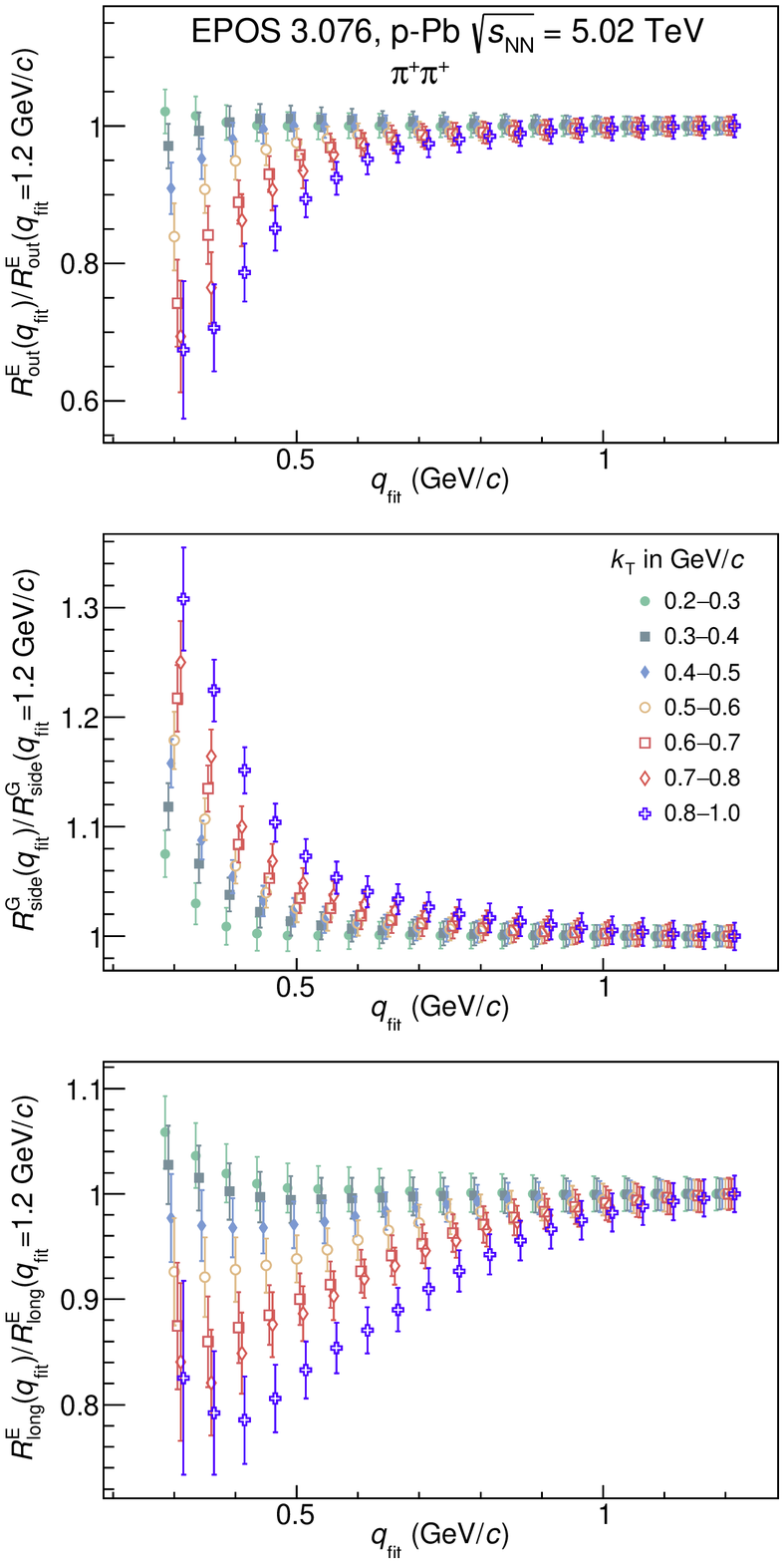}  
\end{minipage}
\caption[]{Dependence of the fitted radii on the maximum fitting range value $q_{\rm fit}$, normalized to the value for the maximum fitting range of $q_{\rm fit}=1.2$~GeV/$c$, in the out (upper panels), side (middle panels), and long (lower panels) directions, for all \kt ranges. The Gaussian fit is shown in the left panels, the EGE in the right panels. Values for different $k_{\rm T}$ ranges are slightly shifted in $q_{\rm fit}$ for visibility.} 
\label{fig:radiiPureCorrTest}
\end{figure}

In Fig.~\ref{fig:radiiPureCorrTest} a full dependence for radii in all directions and in all \kt is shown for both functional forms, normalized to the value obtained for the maximum fitting range $q_{\rm fit}=1.2$~GeV/$c$. All radii, for all $k_{\rm T}$, all directions, and both functional forms, reach a stable value if a sufficiently wide fitting range $q_{\rm fit}$ is selected. However, values for large \kt stabilize for $q_{\rm fit}$ larger by even a factor of 2 than at lower $k_{\rm T}$. This is expected, since the width of the effect grows with \kt (femtoscopic size becomes smaller). In other words, selecting a fixed fitting range for all \kt which is too narrow may introduce an artificial \kt dependence into the fitted radii. Also a narrow fitting range can result in the radii being either lower or higher than the stable value, depending on \kt and the functional form being fitted. Incorrect fitting range selection may result in systematic deviations of up to 30\%. We use a value of the maximum fitting range of 1.0~GeV/$c$, which is enough to obtain stable fitting results for all directions, all $k_{\rm T}$, and both functional forms of the fit.

\section{Fitting the full correlation}
\label{sec:fullfit}

\begin{figure}[!ht]
\centering
\begin{minipage}[!ht][][t]{0.483\linewidth}
\includegraphics*[width=1.0\textwidth]{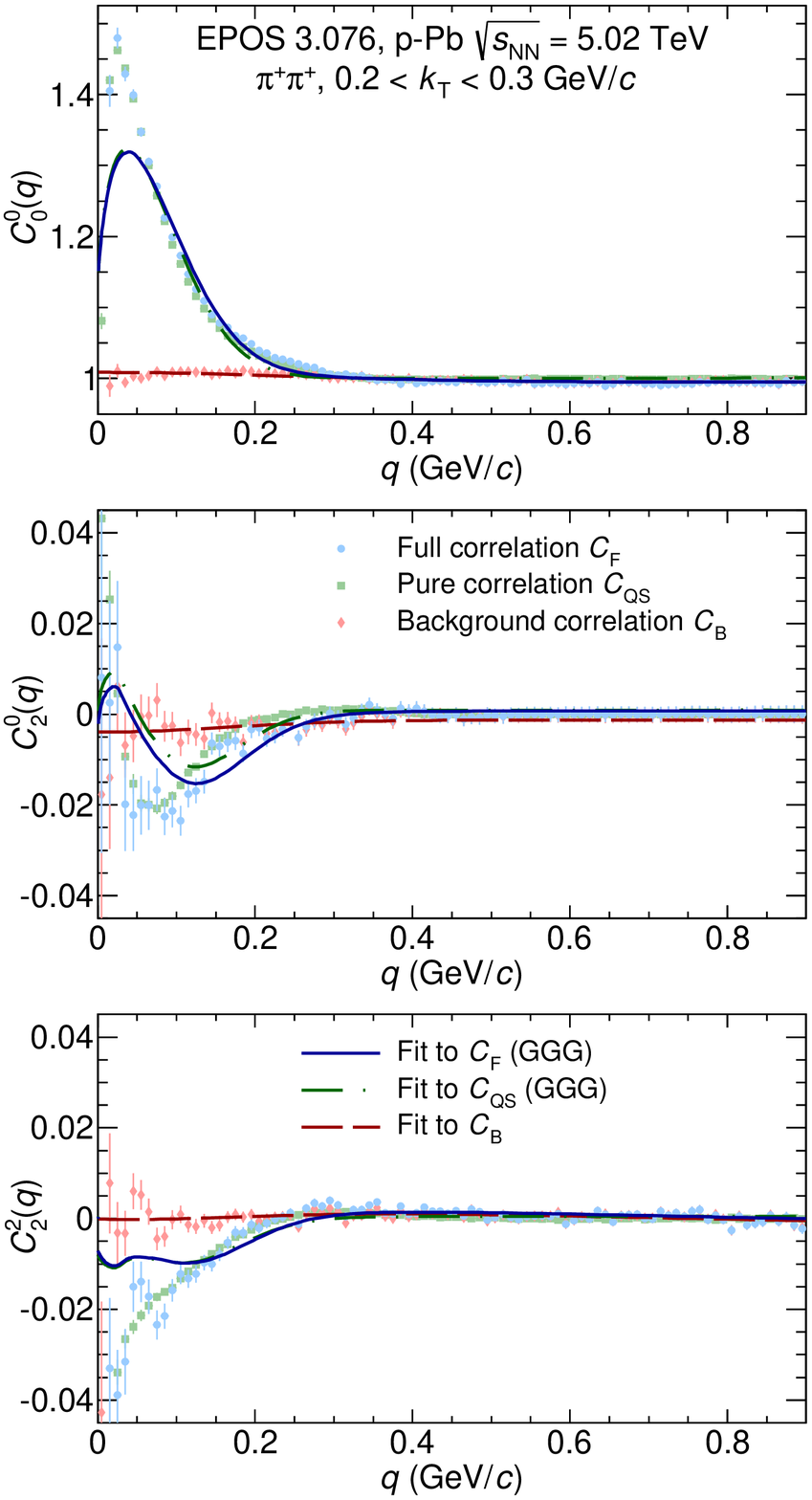} 
\end{minipage}
\hspace{0.25cm}
\begin{minipage}[!ht][][t]{0.483\linewidth}
\includegraphics*[width=1.0\textwidth]{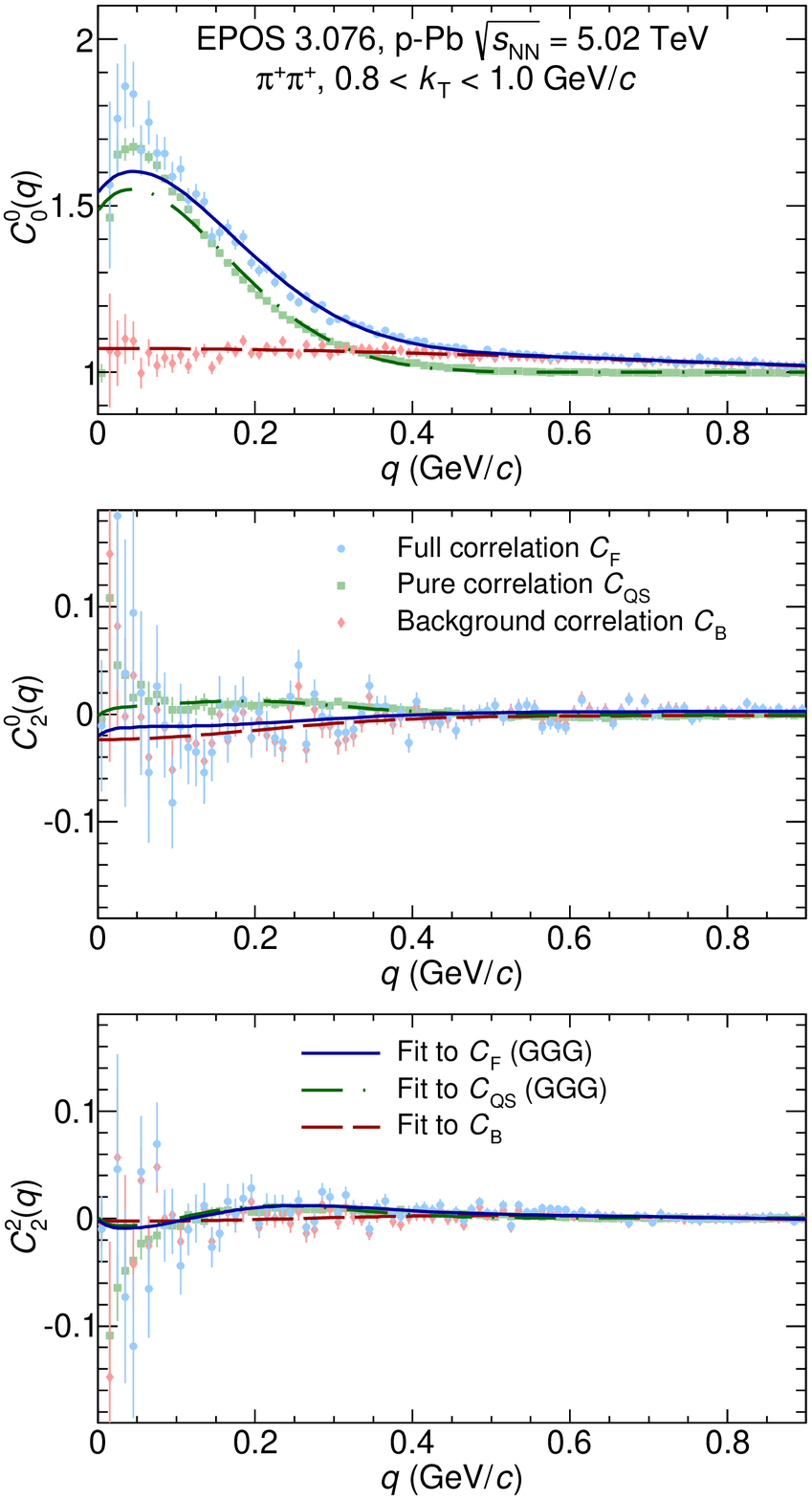}  
\end{minipage}
\caption[]{First three non-vanishing components of the SH representation of the $\pi^+\pi^+$ correlation functions from EPOS model for $0.2<k_{\rm T}<0.3$~GeV/$c$ (left plot) and $0.8<k_{\rm T}<1.0$~GeV/$c$ (right plot). Lines correspond to the GGG fit with maximum fitting range $q_{\rm fit}=1.0$~GeV/$c$.}  
\label{fig:fitsGGG}
\end{figure}

\begin{figure}[!ht]
\centering
\begin{minipage}[!ht][][t]{0.483\linewidth}
\includegraphics*[width=1.0\textwidth]{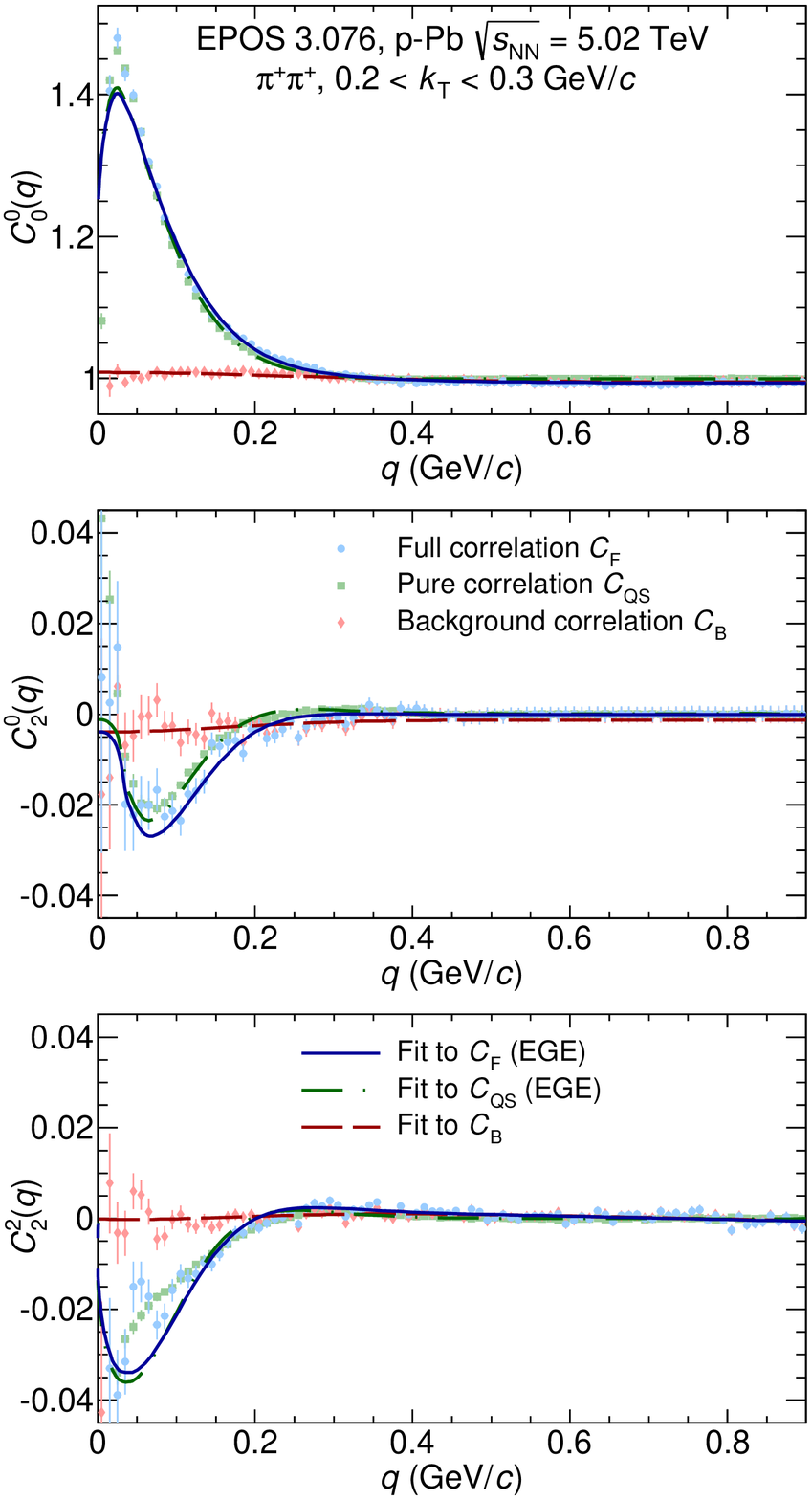} 
\end{minipage}
\hspace{0.25cm}
\begin{minipage}[!ht][][t]{0.483\linewidth}
\includegraphics*[width=1.0\textwidth]{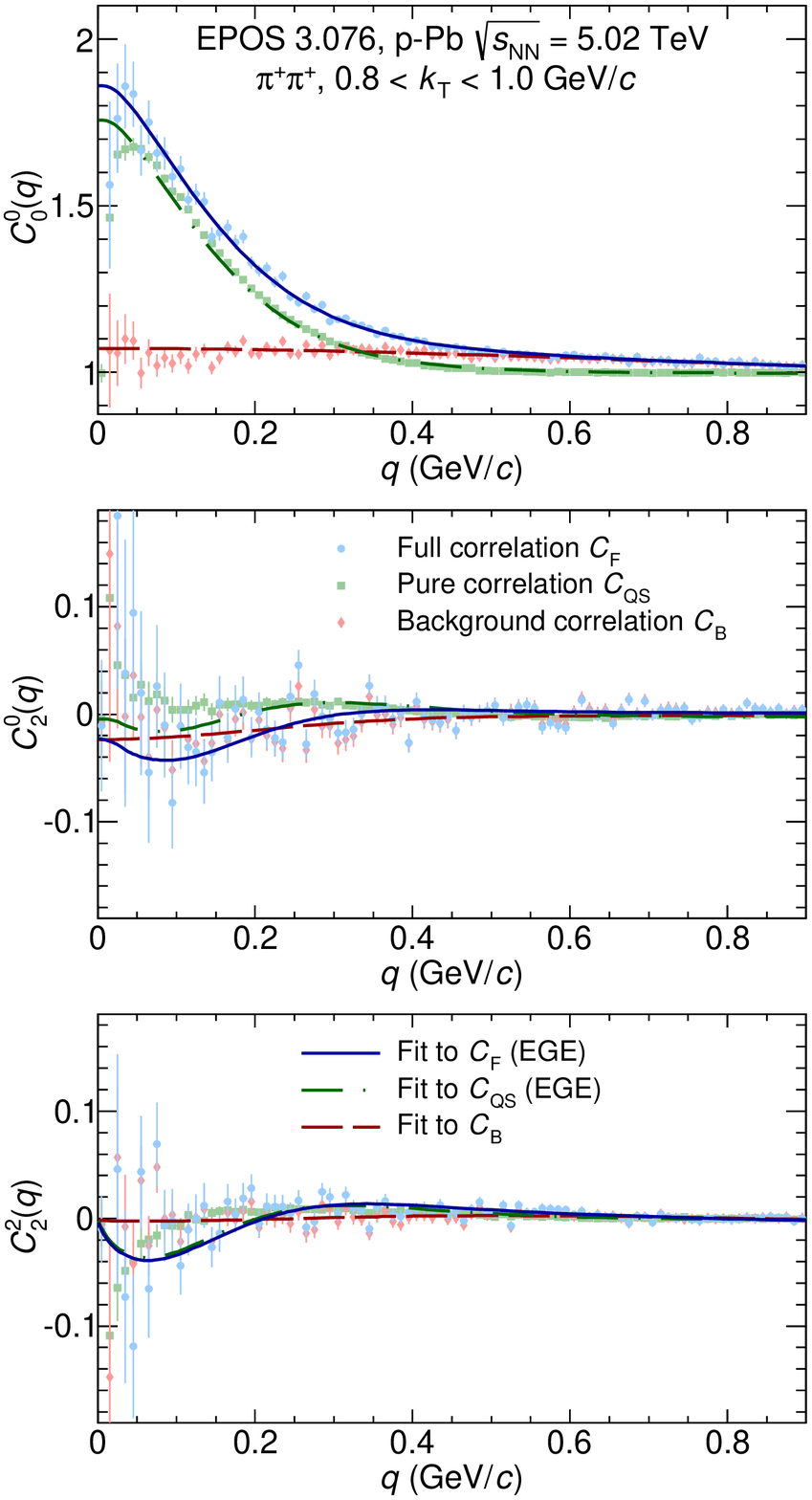}  
\end{minipage}
\caption[]{First three non-vanishing components of the SH representation of the $\pi^+\pi^+$ correlation functions from EPOS model for $0.2<k_{\rm T}<0.3$~GeV/$c$ (left plot) and $0.8<k_{\rm T}<1.0$~GeV/$c$ (right plot). Lines correspond to the EGE fit with maximum fitting range $q_{\rm fit}=1.0$~GeV/$c$.}  
\label{fig:fitsEGE}
\end{figure}

We have performed reference fits to all pure correlation functions $C_{\rm QS}$ for all \kt with the two functional forms. We also proposed three scenarios for the background characterization, which vary in the number of free parameters and the level of Monte Carlo model dependence that they introduce. We now proceed to fit the full correlation function $C_{\rm F}$, which include both the effects of femtoscopic correlations, as well as other event-wide sources. It is now necessary to apply the full fitting formula from Eq.~(\ref{eq:cfitbg}) with the $\Omega$ factor constrained with the procedure described in Section~\ref{sec:back}. Examples of the fits (with maximum fitting range $q_{\rm fit}=1.0$~GeV/$c$) are shown in Fig.~\ref{fig:fitsGGG} for the Gaussian functional form and in Fig.~\ref{fig:fitsEGE} for the EGE fit. The background fit, drawn as red dashed lines in both figures, corresponds to scenario 3, i.e. the full three-dimensional function. It is relatively small for the low $k_{\rm T}$, although even there is some deviation from 1.0 in (0,0) and from 0 in the (2,0) components can be seen. The deviations for the high \kt range are more pronounced. It is also apparent that the Gaussian fit, while able to capture the general trend of the correlation, is not describing the behavior of the correlation function at low $q$. This is fully consistent with experimental observation of non-Gaussian shape of correlation in small system. At the same time the EGE fit works much better in this range, again in agreement with experimental observations. Also the non-trivial behavior of the (2,0) and (2,2) components at low $q$ is better captured by the EGE fit.

\begin{figure}[!ht]
\centering
\begin{minipage}[!ht][][t]{0.49575\linewidth}
\includegraphics*[width=1.0\textwidth]{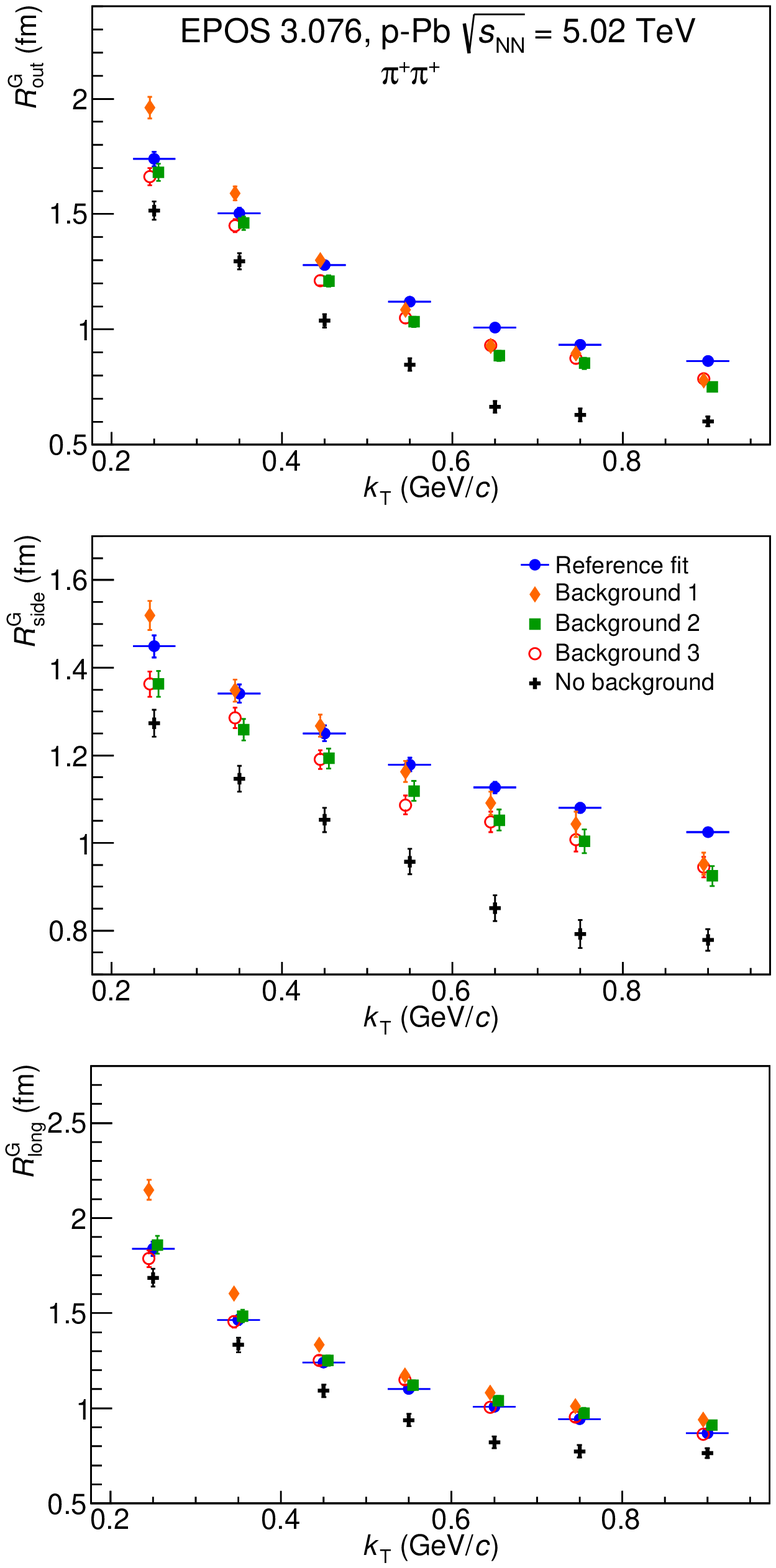} 
\end{minipage}
%\hspace{0.15cm}
\begin{minipage}[!ht][][t]{0.49575\linewidth}
\includegraphics*[width=1.0\textwidth]{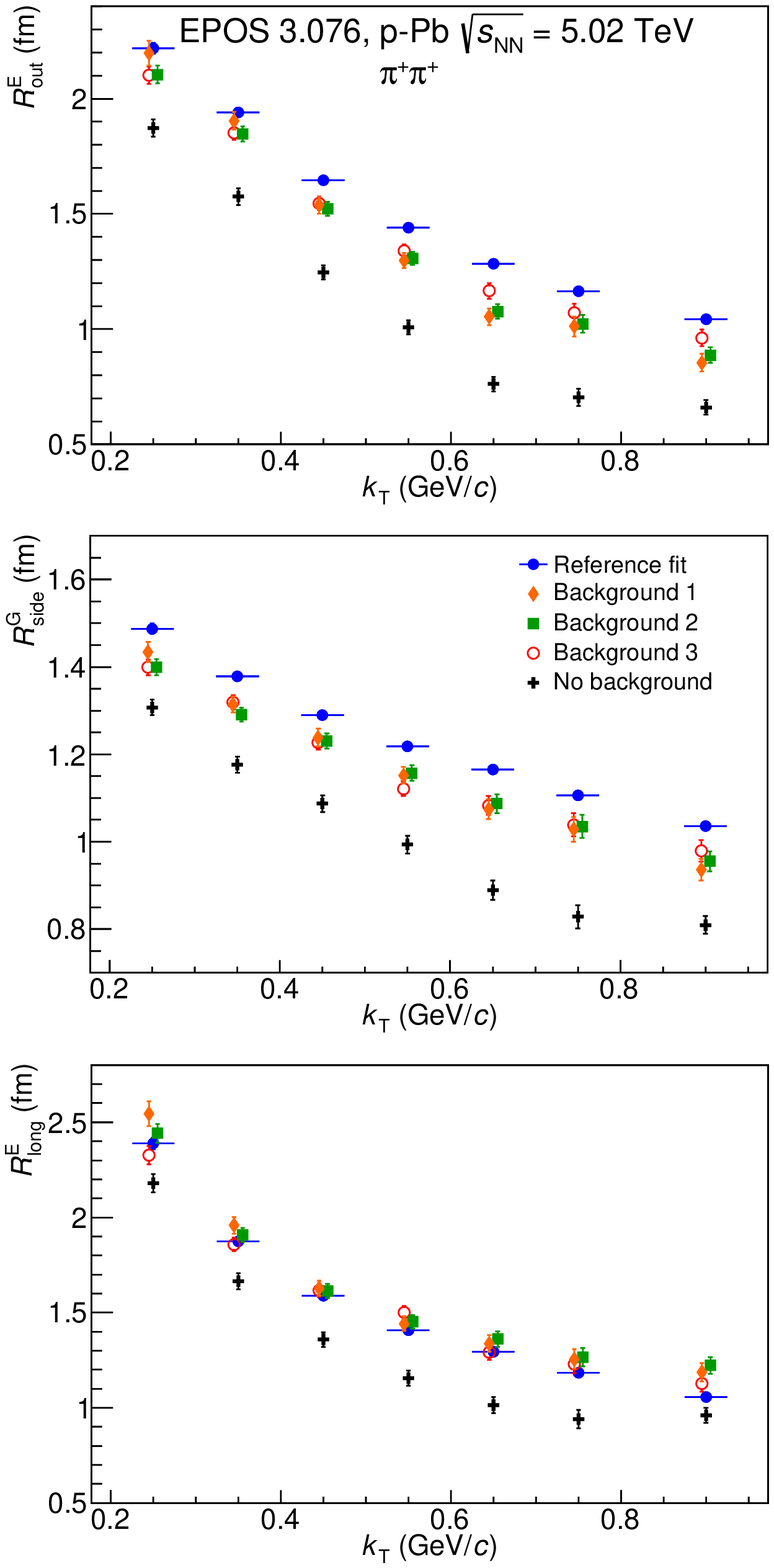}  
\end{minipage}
\caption[]{Extracted femtoscopic radii. "Reference fit" is performed to pure correlation $C_{\rm QS}$. "Background 1": first scenario, with minimal MC dependence (magnitude as free parameter), "Background 2": second scenario, with both magnitude and shape fixed, "Background 3": full three-dimensional shape of the background fixed. "No background": fit to $C_{\rm F}$ is performed with the $\Omega$ factor set to 1.0. All fits are performed with maximum fitting range $q_{\rm fit}=1.0$~GeV/$c$. Points for the same \kt range for various versions of the fit are slightly shifted in \kt for visibility.}  
\label{fig:radii}
\end{figure}

The final radii coming from all the fits are shown in Fig.~\ref{fig:radii}. All three background scenarios are shown, in addition the fit to $C_{\rm F}$ was performed with no background treatment ($\Omega$ in Eq.~(\ref{eq:cfitbg}) set to 1.0). Let us focus first on the extreme case of not accounting for background at all. The radii are then always strongly underpredicted with respect to the reference, with differences reaching 30\%. Such fits are clearly not acceptable in the low multiplicity environment, where significant additional correlation sources are present. All the other scenarios do take the background into account and, as a consequence, they much more closely resemble the reference values. The Gaussian fit with the magnitude of the background free shows relatively large differences. In addition the slope of the \kt dependence is visibly altered -- the radii in the transverse directions are higher than the reference at low \kt and lower at high $k_{\rm T}$. The same fit behaves much better for the EGE case. Nevertheless it seems that trying to constrain the background magnitude with the data itself (leaving the magnitude free in the fit) can potentially dangerously alter the results, unless we precisely control the functional form of the femtoscopic effect. At the moment such form is not known for real collisions, especially in the very fresh p--Pb data at the LHC. Therefore, using a fit with unconstrained background magnitude is also discouraged. That leaves the last two options, where both the magnitude and the shape of the background are constrained based on the Monte Carlo procedure. They both produce comparable agreement with the reference sample, with the full three-dimensional background giving a
slightly better agreement, as should be expected. However, it should be noted that EPOS model produces a relatively spherically symmetric shape of the correlation function, which may not be the case for experimental data. For a Gaussian fit the radii deviate downwards by 4-8\% for the $out$ direction, downward by 6-8\% for the $side$ direction and no more than 3\% in the $long$ direction. For the EGE fit the agreement is very similar. Therefore we have shown that in order to account for the non-femtoscopic effects in the small systems, one needs to first constrain the shape and the magnitude of the background with a Monte Carlo simulation. The remaining systematic uncertainty of the method is then 3-8\%. 

\section{Summary}

We have presented the analysis of the femtoscopic correlation functions for identical pions, calculated for the EPOS model of the p--Pb collisions at $\sqrt{s_{\rm NN}}=5.02$~TeV. Significant non-femtoscopic correlation sources are found to influence such functions, qualitatively consistent with the experimental observations. We propose a robust procedure to account for such correlations in the extraction of the femtoscopic radii. If both the magnitude as well as the shape of the background effects are properly constrained with the help of the Monte Carlo simulation, the correct
values of the radii can be extracted, with systematic uncertainty coming from the method itself equal to approximately 3-8\%. The proper selection of the fitting range was also discussed, and recommendations were given to always use a range that fully includes the femtoscopic signal, together with a reasonable portion of background-dominated region of the relative momentum. 

\begin{appendices}
\appendix
\section{}
\label{appendix}
Moments of the spherical harmonic decomposition of the correlation function are given by:
\begin{equation}
C_{l}^{m}(q)=\frac{1}{\sqrt{4\pi}}\int \mathrm{d}\varphi \mathrm{d}(\cos\theta)C(\mathbf{q}){Y^{m*}_{l}}(\theta,\varphi),
\end{equation}
where $\theta$ and $\varphi$ are the spherical angles, and $Y^{m*}_{l}(\theta,\varphi)=(-1)^mY_l^{-m}(\theta,\varphi)$ are the conjugate spherical harmonic functions, $l$ is a natural number and $m$ is an integer $-l \leq m \leq l$. The components of three-vector $\bf q$ in the LCMS frame are then $q_{\rm long}=|\bf q|\cos\theta$, $q_{\rm out}=|\bf q|\sin\theta\cos\varphi$, and $q_{\rm side}=|\bf q|\sin\theta\sin\varphi$, and in spherical coordinate system: $q=|\mathbf{q}|$, $\theta$, and $\varphi$. In the case of collider experiments and correlations of identical particles, the following components vanish: (1) all imaginary components, (2) odd $l$ components, (3) odd $m$ components for even $l$. The first three non-vanishing components, $C_0^0$, $C_2^0$, and $C_2^2$, capture essentially all the three-dimensional structure of the correlation effect.

The full correlation function $C(\mathbf{q})$ constructed from the spherical harmonic components has therefore the following form:
\begin{equation}
C(\mathbf{q})=\sqrt{4\pi}\sum_{l,m}C_l^m(q)Y_l^m(\varphi,\theta).
\end{equation}
\end{appendices}

The complete formalism of calculation the femtoscopic correlation function in spherical harmonics can be found in Ref.~\cite{Kisiel:2009iw}

\section*{Acknowledgments}
We would like to thank Klaus Werner for providing a p--Pb event sample generated with the EPOS 3.076 code. This work has been financed by the
Polish National Science Centre under decisions no. DEC-2011/01/B/ST2/03483, DEC-2012/05/N/ST2/02757, DEC-2013/08/M/ST2/00598, and by the European Union in the framework of European Social Fund.

\bibliographystyle{ieeetr}
\bibliography{bibliography}
\end{document}